\documentclass[12pt]{article}
\usepackage[latin1]{inputenc}
\usepackage[british]{babel}
\usepackage{cmap}
\usepackage{lmodern}

\usepackage{amssymb, amsmath, amsthm}
\usepackage[a4paper,top=25mm,bottom=25mm,left=25mm,right=25mm]{geometry}
\usepackage{ragged2e}

\usepackage{authblk} % for headings
\usepackage{pifont}
\usepackage{graphicx}
\usepackage[usenames,dvipsnames,svgnames,table]{xcolor}
\usepackage[figuresright]{rotating}
\usepackage{xtab} % tackle the long tables
\usepackage{longtable} % tackle the long tables
\usepackage{multirow}
\usepackage{footnote}
\usepackage[stable]{footmisc}
\usepackage{chngpage} % allows for temporary adjustment of side margins
\usepackage{pdflscape} % landscape environment
\usepackage{tocbibind} % includes everything in ToC

\usepackage{pgfplots}
\pgfplotsset{compat=1.14}
\pgfplotsset{every tick label/.append style={font=\footnotesize}}
\usepackage{setspace}
\makesavenoteenv{tabular}
\usepackage{tabularx}
\usepackage{booktabs}
\usepackage{threeparttable} % [flushleft]
\usepackage[referable]{threeparttablex} % footnotes in tabu
\newcolumntype{R}{>{\raggedleft\arraybackslash}X}
\newcolumntype{L}{>{\raggedright\arraybackslash}X}
\newcolumntype{C}{>{\centering\arraybackslash}X}
\newcolumntype{A}{>{\columncolor{gray!25}}C}
\newcolumntype{a}{>{\columncolor{gray!25}}c}

\newlength{\tablen}

\usepackage{dcolumn} % alignment to decimal points
\newcolumntype{.}{D{.}{.}{-1}}

\usepackage{tikz}
\usetikzlibrary{arrows, calc, matrix, patterns, positioning, trees}
\usepackage[semicolon]{natbib}
\usepackage[hyphens]{url}
\usepackage{hyperref} % [hidelinks]
\hypersetup{
  colorlinks   = true,    		% Colours links instead of ugly boxes
  urlcolor     = blue,    		% Colour for external hyperlinks
  linkcolor    = blue,			% Colour of internal links
  citecolor    = ForestGreen,		  	% Colour of citations
}
\usepackage{microtype}
\usepackage[justification=centering]{caption} % [justification=centering]

\usepackage[labelformat=simple]{subcaption}

\DeclareCaptionLabelFormat{parenthesis}{(#2)}
\makeatletter
\renewcommand\p@subfigure{\arabic{figure}.}
\makeatother

\DeclareCaptionLabelFormat{parenthesis}{(#2)}
\makeatletter
\renewcommand\p@subtable{\arabic{table}.}
\makeatother

\def\addlegendimage{\csname pgfplots@addlegendimage\endcsname}

\usepackage{enumitem}

\setlist[itemize]{leftmargin=2.5\parindent}
\setlist[enumerate]{leftmargin=2.5\parindent}

\theoremstyle{plain}
\theoremstyle{definition}

\newtheorem{example}{Example}%[section]

\theoremstyle{remark}

\def\keywords{\vspace{.5em} % Add keywords
{\noindent \textit{Keywords}: }}

\def\AMS{\vspace{.5em} % Add keywords
{\noindent \textbf{\emph{MSC} class}: }}

\def\JEL{\vspace{.5em} % Add keywords
{\noindent \textbf{\emph{JEL} classification number}: }}

\title{How to avoid uncompetitive games? \\ The importance of tie-breaking rules}
\author{\href{https://sites.google.com/view/laszlocsato}{L\'aszl\'o Csat\'o}\thanks{~E-mail: \emph{laszlo.csato@sztaki.hu}} }
\affil{Institute for Computer Science and Control (SZTAKI) \\
E\"otv\"os Lor\'and Research Network (ELKH) \\
Laboratory on Engineering and Management Intelligence \\
Research Group of Operations Research and Decision Systems}
\affil{Corvinus University of Budapest (BCE) \\
Department of Operations Research and Actuarial Sciences}
\affil{Budapest, Hungary}
\date{\today}

\def\Dedication{
{\noindent
$\mathfrak{Freilich}$ $\mathfrak{kann}$ $\mathfrak{man}$ $\mathfrak{fordern}$, $\mathfrak{da\ss}$ $\mathfrak{dies}$ $\mathfrak{sowenig}$ $\mathfrak{als}$ $\mathfrak{m\ddot{o}glich}$ $\mathfrak{sei}$, $\mathfrak{aber}$ $\mathfrak{nur}$ $\mathfrak{in}$ $\mathfrak{Beziehung}$ $\mathfrak{auf}$ $\mathfrak{den}$ $\mathfrak{einzelnen}$ $\mathfrak{Fall}$: $\mathfrak{n\ddot{a}mlich}$, $\mathfrak{sowenig}$ $\mathfrak{als}$ \underline{$\mathfrak{in}$ $\mathfrak{diesem}$ $\mathfrak{einzelnen}$ $\mathfrak{Fall}$ $\mathfrak{m\ddot{o}glich}$}, $\mathfrak{nicht}$ $\mathfrak{aber}$, $\mathfrak{da\ss}$ $\mathfrak{man}$ $\mathfrak{den}$ $\mathfrak{Fall}$, $\mathfrak{wobei}$ $\mathfrak{die}$ $\mathfrak{Ungewi{\ss}heit}$ $\mathfrak{am}$ $\mathfrak{geringsten}$ $\mathfrak{ist}$, $\mathfrak{immer}$ $\mathfrak{vorziehen}$ $\mathfrak{m\ddot{u}{\ss}te}$; $\mathfrak{das}$ $\mathfrak{w\ddot{a}re}$ $\mathfrak{ein}$ $\mathfrak{ungeheurer}$ $\mathfrak{Versto\ss}$, $\mathfrak{wie}$ $\mathfrak{das}$ $\mathfrak{aus}$ $\mathfrak{allen}$ $\mathfrak{unseren}$ $\mathfrak{theoretischen}$ $\mathfrak{Ansichten}$ $\mathfrak{hervorgehen}$ $\mathfrak{wird}$.\footnote{~
``\emph{Whatever is wanting in certainty must always be left to fate, or chance, call it which you will. We may demand that what is so left should be as little as possible, but only in relation to the particular case---that is, as little as is possible in this one case, but not that the case in which the least is left to chance is always to be preferred. That would be an enormous error, as follows from all our theoretical views.}'' (Source: Carl von Clausewitz: \emph{On War}, Book 2, Chapter 5 [Criticism]. Translated by Colonel James John Graham, London, N. Tr\"ubner, 1873. \url{http://clausewitz.com/readings/OnWar1873/TOC.htm})}
}
\vspace{0.25cm}

\flushright
\noindent (Carl von Clausewitz: \emph{Vom Kriege})

\vspace{1cm} 
\justify }

\begin{document}

\newgeometry{top=10mm,bottom=15mm,left=25mm,right=25mm}
%\newgeometry{top=20mm,bottom=20mm,left=25mm,right=25mm}
\maketitle
\thispagestyle{empty}
\Dedication

\begin{abstract}
\noindent
If the final position of a team is already secured independently of the outcomes of the remaining games in a round-robin tournament, it might play with little enthusiasm. This is detrimental to attendance and can inspire collusion and match-fixing. We demonstrate that tie-breaking rules might affect the occurrence of such a situation. Its probability is quantified via simulations for the four groups of the 2022/23 UEFA Nations League A under two well-established tie-breaking rules, goal difference and head-to-head records. In these home-away round-robin contests with four teams and 12 matches, the competitiveness of the final four games can be promoted by giving priority to goal difference, which reduces the chance of a fixed position in the group ranking by at least two and usually five percentage points in the last round. Our findings, supported by sensitivity analysis in a theoretical model, provide important lessons on how to design ranking systems.


\AMS{62F07, 90-10, 90B90}
% Statistical ranking and selection procedures
% Mathematical modeling or simulation for problems pertaining to operations research and mathematical programming
% Case-oriented studies in operations research

\JEL{C44, C63, Z20}
% Operations Research, Statistical Decision Theory
% Computational Techniques, Simulation Modeling 
% Sports Economics, General
\end{abstract}

\clearpage
\restoregeometry

\section{Introduction} \label{Sec1}

``\emph{Designing an optimal contest is both a matter of significant financial concern for the organizers, participating individuals, and teams, and a matter of consuming personal interest for millions of fans}'' \citep[p.~1137]{Szymanski2003}. One of the most important responsibilities of the administrators is to set the right incentives for the contestants \citep{KendallLenten2017, LentenKendall2021}. It is widely acknowledged that Operational Research (OR) can contribute to tournament design by analysing the effects of policy changes and making proposals to improve the rules \citep{Wright2009, Wright2014, Csato2021a}. 

There exist two fundamental tournament formats \citep{ScarfYusofBilbao2009}.
The \emph{knockout competition} consists of rounds. In each round, the winners progress to the next round and the losers are eliminated. The contest is won by the winner of the final round. Therefore, incentive compatibility is usually not threatened because the players need to win to avoid elimination. Nevertheless, several issues remain to be studied by OR, including fairness \citep{ArlegiDimitrov2020, Arlegi2022} and seeding procedures \citep{HorenRiezman1985, GrohMoldovanuSelaSunde2012, DagaevSuzdaltsev2018}.

In a \emph{round-robin competition}, each competitor plays against all the others and earns points according to its number of wins (and possibly draws). Since the contestants do not face elimination, they may benefit from deliberately losing a game. For example, being ranked second might lead to playing against a preferred competitor in the next round of the tournament \citep{Guyon2022a, Pauly2014, Vong2017}. In certain settings, a team can be strictly better off by losing---not only in expected terms---because qualification is allowed from multiple tournaments \citep{DagaevSonin2018}, teams playing in different round-robin groups are compared \citep{Csato2020c}, or an exogenous ranking of the teams provides a secondary way to qualify \citep{Csato2022a, HaugenKrumer2021}.

However, the last games of a round-robin tournament are sometimes played with little enthusiasm even if the rules are well designed and a high-ranking position is adequately rewarded because the place of a team in the final ranking can already be secured, independently of the results of the remaining matches. Such a team may play below its actual potential, which is detrimental to the integrity of sport and is advantageous for the ``lucky'' opponents that play at the end of the tournament against this particular team. 

Even though a team might exert full effort despite the game being stakeless from its perspective, similar scenarios should be avoided to the extent possible: the mere suspicion of reluctance to invest full effort into winning or using a lower quality squad is still against the spirit of sports. Consequently, the organisers can be blamed for choosing a design that fails to promote competitiveness to the greatest degree.

In particular, the present paper will study the role of tie-breaking rules with respect to the probability that the final position of a team is already known when some matches remain to be played. These often ignored ranking criteria may influence the stakes of a game according to the following illustration.

\begin{table}[ht!]
\begin{threeparttable}
\centering
\caption{Ranking in Group 3 of the 2020/21 UEFA Nations League A \\ before the last matchday}
\label{Table1}
\rowcolors{3}{}{gray!20}
    \begin{tabularx}{\linewidth}{Cl CCC CCC >{\bfseries}C} \toprule
    Pos   & Team  & W     & D     & L     & GF    & GA    & GD    & Pts \\ \bottomrule
    1     & France & 4     & 1     & 0     & 8     & 3     & $+5$    & 13 \\
    2     & Portugal & 3     & 1     & 1     & 9     & 2     & $+7$    & 10 \\
    3     & Croatia & 1     & 0     & 4     & 7     & 13    & $-6$    & 3 \\
    4     & Sweden & 1     & 0     & 4     & 3     & 9     & $-6$    & 3 \\ \bottomrule    
    \end{tabularx}
    
    \begin{tablenotes}
\item
\footnotesize{Pos = Position; W = Won; D = Drawn; L = Lost; GF = Goals for; GA = Goals against; GD = Goal difference; Pts = Points. All teams have played five matches.}   
    \end{tablenotes}
\end{threeparttable}
\end{table}

\begin{example} \label{Examp1}
Table~\ref{Table1} shows the standing of Group 3 in the 2020/21 UEFA Nations League A after five rounds, with the matches Croatia vs.\ Portugal and France vs.\ Sweden still to be played. If two or more teams in the group are equal on points on completion, higher number of points obtained in the matches played among the teams in question decides their position \citep[Article~15.01]{UEFA2020e}. As the result of France vs.\ Portugal has been 0-0, the result of Portugal vs.\ France has been 0-1, and France leads by three points over Portugal, France is guaranteed to win the group and Portugal to be the runner-up. Consequently, there is one team in each of the two matches played in the sixth round whose position cannot change.

On the other hand, if tie-breaking would have been based on goal difference instead of head-to-head results, Portugal could have hoped to be the first with defeating Croatia.
\end{example}

Example~\ref{Examp1} uncovers that the choice of tie-breaking rules can affect the competitiveness of the matches played at the end of the contest.
The previous literature has concentrated primarily on the connection between the schedule (the order) of the games and match-fixing opportunities in round-robin tournaments.
Inspired by the structure of the 2026 FIFA World Cup, \citet{Guyon2020a} examines the risk of collusion in a group of three. Risk of collusion emerges when the two teams playing the only match in the last round can qualify at the expense of the third team. The probability of this scenario is quantified and the schedule minimising its occurrence is identified.
\citet{Stronka2020} investigates ``temptation to lose'' in a group of four with the top two teams qualifying, which results from the desire to play against a weaker opponent in the first round of the subsequent knockout phase. Three pair matching methods are analysed and compared via simulations. Besides changing the pairing algorithm, the schedule also plays a role in decreasing the threat of ``temptation to lose''.
\citet{ChaterArrondelGayantLaslier2021} classify the games played in the last round of the FIFA World Cup into three categories: competitive (neither team is indifferent and they want to achieve incompatible goals), collusive (the targets of the teams are compatible and neither is indifferent), and stakeless (at least one team is completely indifferent between winning, drawing, or losing). The choice of games played in the last round is found to be crucial for making them more exciting to watch.

The study of tie-breaking rules remains more limited.
\citet{Winchester2016} analyses the implications of bonus points used to reward teams for ``coming close'' in losing efforts in most rugby union tournaments. However, this system is not only a tie-breaker as bonus points can lead to situations when teams with fewer wins but more bonus points qualify over teams with more wins.
\citet{Berker2014} evaluates the occurrence rates of \emph{heteronomous relative ranking}---when the relative ranking of two teams depends on the outcome of a match in which neither of them was involved---under the two main tie-breaking principles, goal difference and head-to-head results, which are usually applied by the F\'ed\'eration Internationale de Football Association (FIFA) and the Union of European Football Associations (UEFA), respectively. Head-to-head records exhibit significantly more often this counterintuitive side effect.
According to the arguments of \citet{Pakaslahti2019} on philosophical grounds, tie-breaking in round-robin contests should give more importance to overall goal difference than to head-to-head results.
\citet[Chapter~1.3]{Csato2021a} reveals the lack of consensus concerning tie-breaking criteria in the top-tier association football (henceforth football) leagues across Europe.

The novelty of the current research resides in the analysis of some tie-breaking rules used in round-robin tournaments from an innovative perspective, namely, the collusion opportunities created in the matches played in the last round(s).
Our main contributions can be summarised as follows:
\begin{itemize}
\item
The role of two well-established tie-breaking criteria, goal difference and head-to-head records, in promoting competitiveness is explored;
\item
A relatively simple but efficient methodology is presented to identify situations where a team has few incentives to exert full effort and to compute the probability of reaching them;
\item
The four groups of the 2022/23 UEFA Nations League A are compared with respect to the threat of stakeless games under the two basic tie-breaking principles. The most important differences between the two rules seem to be independent of the distribution of teams' strengths.
\end{itemize}
In addition, it is worth noting that the tournament considered here---four teams playing in a home-away round-robin format with 12 games---is more difficult to analyse than the ones appearing in the literature since \citet{Guyon2020a} deals with the case of three teams playing a single round-robin with three matches, while \citet{Stronka2020} and \citet{ChaterArrondelGayantLaslier2021} examine single round-robin with four teams and six matches. Therefore, we should account for more possible scenarios.

On the other hand, the major implication is in line with the literature \citep{Berker2014, Pakaslahti2019}: the priority of head-to-head results over goal difference may more often lead to unfavourable situations, thus, goal difference is a better tie-breaking rule compared to head-to-head records. Our findings provide an essential lesson for tournament organisers on how to design ranking systems.

Last but not least, it needs to be emphasised that there are other---albeit less widely used---tie-breaking rules applied in practice.\footnote{~We are grateful to an anonymous referee for calling our attention to this limitation of the research.}
The rugby union bonus points system has already been mentioned, although it is not only a tie-breaking principle \citep{Winchester2016}. In certain top-tier football leagues, the number of wins is the primary tie-breaking criterion \citep[Chapter~1.3]{Csato2021a}. Goal average or goal ratio (the number of goals scored divided by the number of goals conceded) was the original tie-breaking rule in football, and is still used in Australian rules football under the name ``percentage''.

The paper is structured as follows. Section~\ref{Sec2} presents the background of the simulation experiment. In particular, the 2022/23 UEFA Nations League A is outlined in Section~\ref{Sec21}, the simulation model is described in Section~\ref{Sec22}, and the two tie-breaking options are defined in Section~\ref{Sec23}. Section~\ref{Sec3} contains the main results: Section~\ref{Sec31} determines the set of matches for which the outcome does not affect the position of a team, Section~\ref{Sec32} overviews the simulation procedure, while Sections~\ref{Sec33} and \ref{Sec34} investigate the two popular tie-breaking principles in the 2022/23 UEFA Nations League A and in a basic theoretical model, respectively. Finally, Section~\ref{Sec4} offers concluding remarks.

\section{Methodology} \label{Sec2}

In the following, the basics of the quantitative evaluation are detailed to allow its replication.

\subsection{The format of the 2022/23 UEFA Nations League A} \label{Sec21}

The 2022/23 UEFA Nations League is the third season of the UEFA Nations League, an international association football competition contested by men's national teams. The 55 UEFA member associations are divided into four leagues. In the top division called League A, the 16 teams play in four home-away round-robin groups of four teams each.

\begin{table}[t!]
  \centering
  \caption{The 2022/23 UEFA Nations League A}
  \label{Table2} 
\rowcolors{3}{gray!20}{} 
    \begin{tabularx}{0.7\textwidth}{Lc m{1cm} Lc} \toprule \hiderowcolors
        \multicolumn{2}{c}{\textbf{Group 1}} &       & \multicolumn{2}{c}{\textbf{Group 2}} \\ %\midrule
    Team 	& Elo   &       & Team & Elo \\ \bottomrule \showrowcolors
    France  & 2114  &       & Spain & 2037 \\
    Denmark & 1937  &       & Portugal & 1972 \\
    Croatia & 1858  &       & Switzerland & 1934 \\
    Austria & 1731  &       & Czech Republic & 1833 \\ \bottomrule
	\end{tabularx}

\vspace{0.25cm}
\begin{threeparttable}
\rowcolors{3}{gray!20}{}	
    \begin{tabularx}{0.7\textwidth}{Lc m{1cm} Lc} \toprule \hiderowcolors
    \multicolumn{2}{c}{\textbf{Group 3}} &       & \multicolumn{2}{c}{\textbf{Group 4}} \\
    Team	& Elo   &       & Team & Elo \\ \bottomrule \showrowcolors
    Italy   & 2030  &       & Belgium & 2075 \\
    Germany & 1963  &       & Netherlands & 1929 \\
    England & 2032  &       & Poland & 1770 \\
    Hungary & 1726  &       & Wales & 1836 \\ \bottomrule
    \end{tabularx}
\begin{tablenotes} \footnotesize
\item
The strengths of the teams are measured by their World Football Elo Ratings on 16 December 2021 (the date of the group draw), see \url{https://www.international-football.net/elo-ratings-table?year=2021&month=12&day=16&confed=UEFA}.
\end{tablenotes}
\end{threeparttable}
\end{table}

The composition of the groups is presented in Table~\ref{Table2}.
The four group winners advance to the 2023 UEFA Nations League Finals and have a chance to become the UEFA Nations League champions. The fourth-placed team in each group is relegated to the 2024/25 UEFA Nations League B. The seeding of the 2024/25 UEFA Nations League A is based on the results of the 2022/23 UEFA Nations League A: the group winners are drawn from Pot 1, the runners-up are drawn from Pot 2, and the third-placed teams are drawn from Pot 3. Therefore, it is reasonable to assume that a team exerts full effort if it can be ranked higher with a better result but it plays with little enthusiasm if its position in the final group ranking is already known. The organiser aims to avoid the latter situation to the extent possible.

\subsection{Simulating match outcomes} \label{Sec22}

Historical tournament data can provide at most limited conclusions since they represent only a single realisation of several random variables \citep{ScarfYusofBilbao2009}. Therefore, we have chosen to use simulations, a standard methodology for analysing and comparing tournament designs \citep{ChaterArrondelGayantLaslier2021, DagaevRudyak2019, GoossensBelienSpieksma2012, LasekGagolewski2018}.

In order to predict the outcomes of individual ties, the strengths of the teams should be quantified. Even though the FIFA World Ranking has been revised in 2018 \citep{FIFA2018c}, the new formula has still some shortcomings such as the lack of home advantage and the missing adjustment for goal difference. Both factors are addressed by the World Football Elo Ratings, available at the website \href{http://eloratings.net/}{eloratings.net}, which has been widely used in scientific research \citep{CeaDuranGuajardoSureSiebertZamorano2020, GasquezRoyuela2016, HvattumArntzen2010, LasekSzlavikBhulai2013, LasekSzlavikGagolewskiBhulai2016}. The Elo ratings of the teams participating in the 2022/23 UEFA Nations League A are shown---on the day of the group draw---in Table~\ref{Table2}.

In a given match, the numbers of goals scored by the two teams need to be specified. For this purpose, the traditional Poisson model is used \citep{Maher1982, LeyVandeWieleVanEeetvelde2019, VanEetveldeLey2019}. In particular, the probability that team $i$ scores $k$ goals against team $j$ on field $f$ is
\begin{equation} \label{Poisson_dist}
P_{ij}(k) = \frac{ \left( \lambda_{ij}^{(f)} \right)^k \exp \left( -\lambda_{ij}^{(f)} \right)}{k!},
\end{equation}
where $\lambda_{ij}^{(f)}$ is the expected number of goals scored by team $i$ against team $j$ if the match is played on field $f$, which is either home ($f = h$) or away ($f = a$).

The official formula of the World Football Elo Ratings provides win expectancy as follows:
\[
W_{ij} = \frac{1}{1 + 10^{-(E_i + 100 - E_j)/400}},
\]
with $E_i$ and $E_j$ being the Elo ratings of teams $i$ and $j$, respectively. Note that the home advantage is fixed at 100 points.

\citet{FootballRankings2020} has estimated the parameter $\lambda_{ij}^{(f)}$ by a quartic polynomial of the win expectancy $W_{ij}$ using a least squares regression with a regime change based on more than 29 thousand matches played by national football teams.
The expected number of goals for the home team $i$ equals
\begin{equation} \label{Exp_goals_home}
\lambda_{ij}^{(h)} = 
\left\{ \begin{array}{ll}
-5.42301 \cdot W_{ij}^4 + 15.49728 \cdot W_{ij}^3 \\
- 12.6499 \cdot W_{ij}^2 + 5.36198 \cdot W_{ij} + 0.22862 & \textrm{if } W_{ij} \leq 0.9 \\ \\
231098.16153 \cdot (W_{ij}-0.9)^4 - 30953.10199 \cdot (W_{ij}-0.9)^3 & \\
+ 1347.51495 \cdot (W_{ij}-0.9)^2 - 1.63074 \cdot (W_{ij}-0.9) + 2.54747 & \textrm{if } W_{ij} > 0.9,
\end{array} \right.
\end{equation}
and the average number of goals for the away team $j$ is
\begin{equation} \label{Exp_goals_away}
\lambda_{ij}^{(a)} = 
\left\{ \begin{array}{ll}
90173.57949 \cdot (W_{ij} - 0.1)^4 + 10064.38612 \cdot (W_{ij} - 0.1)^3 \\
+ 218.6628 \cdot (W_{ij} - 0.1)^2 - 11.06198 \cdot (W_{ij} - 0.1) + 2.28291 & \textrm{if } W_{ij} < 0.1 \\ \\
-1.25010 \cdot W_{ij}^4 -  1.99984 \cdot W_{ij}^3 & \\
+ 6.54946 \cdot W_{ij}^2 - 5.83979 \cdot W_{ij} + 2.80352 & \textrm{if } W_{ij} \geq 0.1.
\end{array} \right.
\end{equation}

%This model has recently been adopted to compute the probability of qualification for the 2018 FIFA World Cup \citep{Csato2022c} and to quantify the incentive incompatibility of the European Qualifiers for the 2022 FIFA World Cup \citep{Csato2022a}.

\subsection{Alternative ranking rules} \label{Sec23}

If two or more teams in the same group collect the same number of points, their order should be decided by tie-breaking rules. There are two basic concepts for this purpose: \emph{head-to-head records} and \emph{goal difference}. UEFA usually gives priority to head-to-head results, which also holds for the 2022/23 UEFA Nations League \citep[Article~15]{UEFA2021h}.

In our model, the \emph{UEFA rule} is defined as follows.
The ranking of teams with the same number of points is determined according to the criteria below:
\begin{enumerate}[label=\alph*)]
\item \label{item1}
higher number of points obtained in the group matches played among the teams in question;
\item
superior goal difference from the group matches played among the teams in question;
\item \label{item3}
higher number of goals scored in the group matches played among the teams in question;
\item
if, after having applied criteria \ref{item1} to \ref{item3}, teams still have an equal ranking, criteria \ref{item1} to \ref{item3} are reapplied exclusively to the matches between the remaining teams to determine their final rankings. If this procedure does not lead to a decision, criteria \ref{item5} to \ref{item7} apply in the order given to the two or more teams still equal;
\item \label{item5}
superior goal difference in all group matches;
\item
higher number of goals scored in all group matches;
\item \label{item7}
drawing of lots.
\end{enumerate}
To summarise, first head-to-head records (if necessary, in a recursive manner), then overall goal difference and the number of goals scored are used to break the ties.

On the other hand, FIFA gives priority to goal difference, see, for instance, the rules of the 2022 FIFA World Cup qualification tournaments \citep[Article~20.6]{FIFA2021}. Therefore, the \emph{FIFA rule} is defined as follows.
To determine the ranking of teams with the same number of points, the criteria below are applied:
\begin{enumerate}[label=\alph*)]
\item
superior goal difference in all group matches;
\item
higher number of goals scored in all group matches;
\item
higher number of points obtained in the group matches played among the teams in question;
\item
superior goal difference from the group matches played among the teams in question;
\item
higher number of goals scored in the group matches played among the teams in question;
\item
drawing of lots.
\end{enumerate}
This tie-breaking rule is based on overall goal difference and the number of goals scored, followed by head-to-head records (without recursion since FIFA does not apply it).

The FIFA and UEFA ranking rules differ in the order of the tie-breaking criteria. It will turn out that the seemingly irrelevant choice has non-negligible sporting effects.

\section{The comparison of tie-breaking options} \label{Sec3}

This section identifies the situations where the position of a team is already secured in the final group ranking and estimates their probabilities of occurrence for the 2022/23 UEFA Nations League A. The two tie-breaking options are examined in a theoretical model, too.

\subsection{The threshold rules} \label{Sec31}

As has been presented in the Introduction, sometimes the place of a team in the final ranking cannot change when some matches are still to be played. Furthermore, its position can be fixed only under the UEFA rule, while this is not the case when the FIFA rule is used to break the ties. The probability of having a fixed position before all games are played will be determined as follows. 

In a home-away round-robin group with four teams, each team plays six matches. Therefore, the first point where the position of a team can already be secure is before Round 5, after four rounds have been played. In particular:
\begin{itemize}
\item
The group winner is fixed under both the FIFA and UEFA ranking rules if it has at least seven points more than the runner-up;
\item
The group winner is fixed under the UEFA ranking rule if
\begin{itemize}[label=$\diamond$]
\item
it has six points more than the runner-up; and
\item
it has at least seven points more than the third-placed team; and
\item
it has played two matches against the runner-up.\footnote{~Technically, we check an equivalent condition. The results of the four matches played in the last two rounds are assumed to be 0-0. The group winner is fixed only under the UEFA rule after four matchdays if and only if the first team has 14, the second team has 8, and the third team has at most 7 points such that the first and the second teams do not play against each other in the last two rounds.}
\end{itemize}
\item
The last team cannot be fixed under the FIFA ranking rule;
\item
The last team is fixed under the UEFA ranking rule if
\begin{itemize}[label=$\diamond$]
\item
it has six points less than the third-placed team; and
\item
it has at least seven points less than the runner-up; and
\item
it has played two matches against the third-placed team.\footnote{~Technically, we check an equivalent condition. The results of the four matches played in the last two rounds are assumed to be 0-0. The last team is fixed under the UEFA rule after four matchdays if and only if the fourth team has 2, the third team has 8, and the second team has at least 9 points such that the fourth and the third teams do not play against each other in the last two rounds.}
\end{itemize}
\end{itemize}

Before Round 6, the possible cases are cumbersome to describe by analogous criteria. But they can be found by analysing what would be the group ranking under the given rule in all extreme cases. In particular, the results of the two remaining matches are assumed to be: (a) M-0, M-0; (b) M-0, 0-M; (c) 0-M, M-0; and (d) 0-M, 0-M with M being a high number.\footnote{~In our computer code, M equals 1000 since it is reasonable to assume that the goal difference of any team will be the highest/lowest if it wins/loses by 1000 goals in the last round.}
The position of a team is secure under the FIFA/UEFA rule if it is the same for all outcomes (a) to (d).

%Finally, it is also interesting to know whether a team whose position is secured only by the UEFA rule obtains the same place under the FIFA rule in the final ranking or not. This can be checked by simulating all group matches.

\begin{table}[t!]
  \centering
  \caption{Matches in the last two rounds of the 2022/23 UEFA Nations League A}
  \label{Table3} 
\centerline{
\rowcolors{3}{gray!20}{} 
    \begin{tabularx}{1.1\textwidth}{cLL c cLL} \toprule \hiderowcolors
    \multicolumn{3}{c}{\textbf{Group 1}} &       & \multicolumn{3}{c}{\textbf{Group 2}} \\
    Round & Home team & Away team &       & Round & Home team & Away team \\ \bottomrule \showrowcolors
    5     & France & Austria &       & 5     & Spain & Switzerland \\
    5     & Croatia & Denmark &       & 5     & Czech Republic & Portugal \\ \hline
    6     & Denmark & France &       & 6     & Portugal & Spain \\
    6     & Austria & Croatia &       & 6     & Switzerland & Czech Republic \\ \bottomrule
	\end{tabularx}
}

\vspace{0.25cm}

\centerline{
\rowcolors{3}{gray!20}{}	
    \begin{tabularx}{1.1\textwidth}{cLL c cLL} \toprule \hiderowcolors
    \multicolumn{3}{c}{\textbf{Group 3}} &       & \multicolumn{3}{c}{\textbf{Group 4}} \\
    Round & Home team & Away team &       & Round & Home team & Away team \\ \bottomrule \showrowcolors
    5     & Italy & England &       & 5     & Belgium & Wales \\
    5     & Germany & Hungary &       & 5     & Poland & Netherlands \\ \hline
    6     & Hungary & Italy &       & 6     & Netherlands & Belgium \\
    6     & England & Germany &       & 6     & Wales & Poland \\ \bottomrule
    \end{tabularx}
}
\end{table}

Obviously, the schedule of the group matches influences the occurrence of these situations.
The last two rounds of matches in the 2022/23 UEFA Nations League A, played in September 2022, are presented in Table~\ref{Table3}.

\subsection{An overview of the simulation exercise} \label{Sec32}

Now all components are available to perform the simulation, which consists of the following steps:
\begin{enumerate}
\item
Setting the input data: the strengths of the teams as measured by the World Football Elo Ratings (Table~\ref{Table2}) and the schedule of the matches (Table~\ref{Table3}); 
\item
Determining the outcome of all matches played in the home-away round-robin tournament (the format of the 2022/23 UEFA Nations League groups, see Section~\ref{Sec21}) based on the Poisson model described in Section~\ref{Sec22}, where the parameters for the expected number of goals are obtained from an external source \citep{FootballRankings2020};
\item
Calculating the proportion of fixed positions before Rounds 5 and 6 under FIFA and UEFA ranking criteria (Section~\ref{Sec23}) according to the threshold rules given in Section~\ref{Sec31}. Naturally, the results of the games played on the last matchday(s) are not taken into account but they are also simulated to know whether a team whose position is fixed only by the UEFA rule obtains the same place under the FIFA rule in the final ranking or not.
\end{enumerate}
One million simulation runs have been performed for each set of inputs.

\subsection{Computational results for the 2022/23 UEFA Nations League} \label{Sec33}

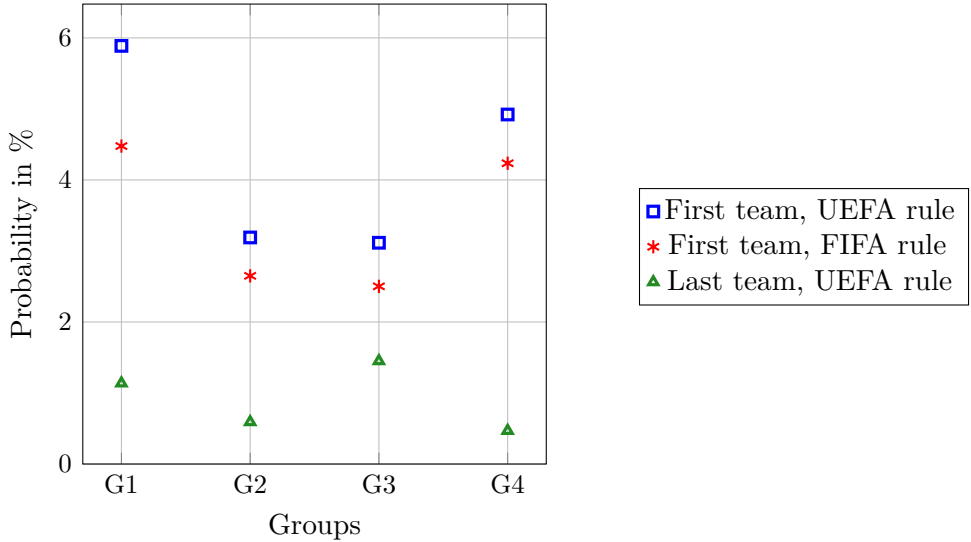
\begin{figure}[t!]
\centering

\begin{tikzpicture}
\begin{axis}[
width = 0.48\textwidth, 
height = 0.48\textwidth,
%title = Unweighted rounds,
%title style = {font=\small},
xmajorgrids = true,
ymajorgrids = true,
symbolic x coords = {G1,G2,G3,G4},
xtick = data,
xlabel = {Groups},
xlabel style = {align=center, font=\small},
ymin = 0,
scaled y ticks = false,
ylabel = {Probability in \%},
ylabel style = {align=center, font=\small},
yticklabel style = {/pgf/number format/fixed,/pgf/number format/precision=5},
ytick style = {draw = none},
legend style = {font=\small,at={(1.2,0.6)},anchor=north west,legend columns=1},
legend entries = {{First team, UEFA rule}, {First team, FIFA rule}$\,$, {Last team, UEFA rule}}
]
% First team UEFA
\addplot [blue, only marks, mark = square, very thick] coordinates{
(G1,5.8874)
(G2,3.1901)
(G3,3.1154)
(G4,4.9213)
};
% First team FIFA
\addplot [red, only marks, mark = asterisk, thick, mark size=2.5pt] coordinates{
(G1,4.477)
(G2,2.6496)
(G3,2.5017)
(G4,4.2349)
};
% Last team UEFA
\addplot [ForestGreen, only marks, mark = triangle, very thick] coordinates{
(G1,1.1373)
(G2,0.5916)
(G3,1.4503)
(G4,0.4687)
};
\end{axis}
\end{tikzpicture}

\captionsetup{justification=centering}
\caption{The probability of an already secured position \\ before Round 5, 2022/23 UEFA Nations League A}
\label{Fig1}

\end{figure}

%\end{document}

Figure~\ref{Fig1} plots the probability that the position of a team is secured after four rounds in the 2022/23 UEFA Nations League A. The group winner is known with a chance of more than 3\% under the UEFA rule, but this value is decreased by at least 50 basis points (0.5\%) under the FIFA rule. The first team will be fixed with the highest probability in Groups 1 and 4, where the difference between the strengths of the best and the second best teams is the highest (see Table~\ref{Table2}). The fourth-placed team can be known only if the UEFA rule is used, the corresponding probability always exceeds 0.4\% and can be close to 1.5\% in Group 3, which contains one weak team, Hungary.

It might be argued that the difference between the UEFA and FIFA ranking rules is overwhelmingly theoretical as a team whose position is known by the UEFA rule will be the first (fourth) at the end even if the FIFA rule is followed. Therefore, we have calculated the associated conditional probability of having a different final position under the FIFA rule if it is already known after four rounds under the UEFA rule. These are ranged between 0.16\% (Groups 1 and 4) and 0.56\% (Group 2) for the group winner, and between 0.19\% (Group 3) and 0.32\% (Groups 2 and 4) for the team to be relegated. Consequently, the advantage of the FIFA regulation over the UEFA is moderated since there are few scenarios where a team should exert full effort in its last two matches only due to the priority given to goal difference.

\begin{figure}[t!]
\centering

\begin{tikzpicture}
%\selectcolormodel{gray}
\begin{axis}[
name = axis1,
width = 0.48\textwidth, 
height = 0.48\textwidth,
title = {Group winner},
title style = {align=center, font=\small},
xmajorgrids = true,
ymajorgrids = true,
symbolic x coords = {G1,G2,G3,G4},
xtick = data,
xlabel = {Groups},
xlabel style = {align=center, font=\small},
ymin = 0,
ymax = 55,
ytick distance = 10,
scaled y ticks = false,
ytick distance = {10},
ylabel = {Probability in \%},
ylabel style = {align=center, font=\small},
yticklabel style = {/pgf/number format/fixed,/pgf/number format/precision=5},
ytick style = {draw = none},
]
% UEFA rule
\addplot [blue, only marks, mark = square, very thick] coordinates{
(G1,46.8535)
(G2,27.3948)
(G3,23.1302)
(G4,41.9484)
};
% FIFA rule
\addplot [red, only marks, mark = asterisk, thick, mark size=2.5pt] coordinates{
(G1,42.9219)
(G2,23.3255)
(G3,15.3371)
(G4,38.2129)
};
\end{axis}

\begin{axis}[
at = {(axis1.south east)},
xshift = 0.15\textwidth,
width = 0.48\textwidth, 
height = 0.48\textwidth,
title = {Runner-up},
title style = {align=center, font=\small},
xmajorgrids = true,
ymajorgrids = true,
symbolic x coords = {G1,G2,G3,G4},
xtick = data,
xlabel = {Groups},
xlabel style = {align=center, font=\small},
ymin = 0,
ymax = 26,
ytick distance = 5,
scaled y ticks = false,
ylabel = {Probability in \%},
ylabel style = {align=center, font=\small},
yticklabel style = {/pgf/number format/fixed,/pgf/number format/precision=5},
ytick style = {draw = none},
]
% UEFA rule
\addplot [blue, only marks, mark = square, very thick] coordinates{
(G1,19.2037)
(G2,16.068)
(G3,16.6519)
(G4,17.7066)
};
% FIFA rule
\addplot [red, only marks, mark = asterisk, thick, mark size=2.5pt] coordinates{
(G1,12.9145)
(G2,11.7625)
(G3,11.5912)
(G4,12.7048)
};
\end{axis}
\end{tikzpicture}

\vspace{0.2cm}
\begin{tikzpicture}
\begin{axis}[
name = axis3,
width = 0.48\textwidth, 
height = 0.48\textwidth,
title = {Third-placed team},
title style = {align=center, font=\small},
xmajorgrids = true,
ymajorgrids = true,
symbolic x coords = {G1,G2,G3,G4},
xtick = data,
xlabel = {Groups},
xlabel style = {align=center, font=\small},
ymin = 0,
ymax = 26,
ytick distance = 5,
scaled y ticks = false,
ylabel = {Probability in \%},
ylabel style = {align=center, font=\small},
yticklabel style = {/pgf/number format/fixed,/pgf/number format/precision=5},
ytick style = {draw = none},
legend style = {font=\small,at={(0.65,-0.25)},anchor=north west,legend columns=4},
legend entries = {UEFA rule$\qquad \qquad \qquad$, FIFA rule}
]
% UEFA rule
\addplot [blue, only marks, mark = square, very thick] coordinates{
(G1,17.7652)
(G2,16.3415)
(G3,18.3776)
(G4,21.2265)
};
% FIFA rule
\addplot [red, only marks, mark = asterisk, thick, mark size=2.5pt] coordinates{
(G1,12.025)
(G2,12.4801)
(G3,13.0782)
(G4,17.4908)
};
\end{axis}

\begin{axis}[
at = {(axis3.south east)},
xshift = 0.15\textwidth,
width = 0.48\textwidth, 
height = 0.48\textwidth,
title = {Fourth-placed team},
title style = {align=center, font=\small},
xmajorgrids = true,
ymajorgrids = true,
symbolic x coords = {G1,G2,G3,G4},
xtick = data,
xlabel = {Groups},
xlabel style = {align=center, font=\small},
ymin = 0,
ymax = 55,
ytick distance = 10,
scaled y ticks = false,
ylabel = {Probability in \%},
ylabel style = {align=center, font=\small},
yticklabel style = {/pgf/number format/fixed,/pgf/number format/precision=5},
ytick style = {draw = none},
]
% UEFA rule
\addplot [blue, only marks, mark = square, very thick] coordinates{
(G1,37.4949)
(G2,21.2468)
(G3,51.7287)
(G4,19.2365)
};
% FIFA rule
\addplot [red, only marks, mark = asterisk, thick, mark size=2.5pt] coordinates{
(G1,33.7167)
(G2,17.1902)
(G3,44.8155)
(G4,15.7118)
};
\end{axis}
\end{tikzpicture}

\caption{The probability of an already secured position \\ before Round 6, 2022/23 UEFA Nations League A}
\label{Fig2}

\end{figure}
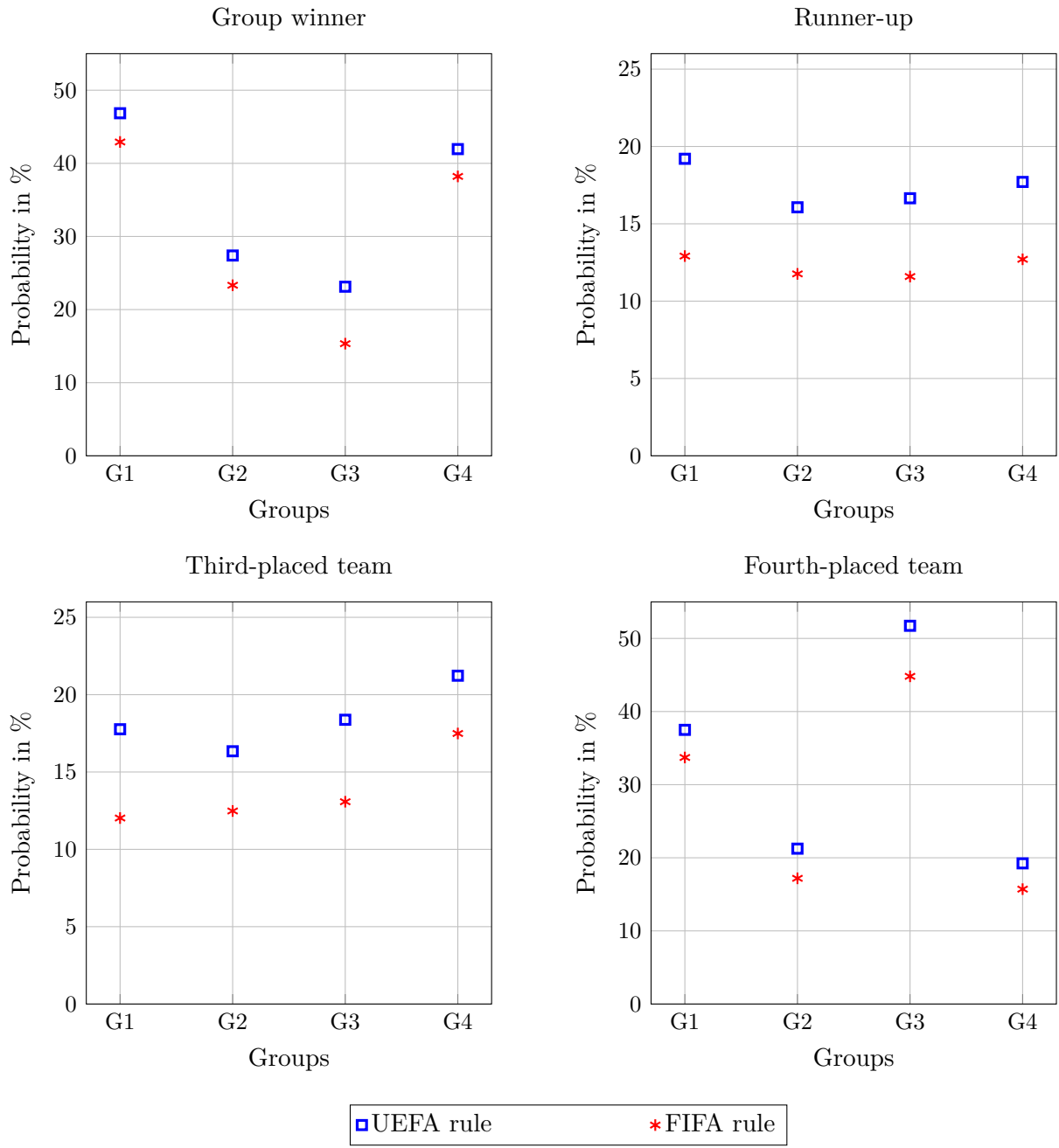

%\end{document}

Figure~\ref{Fig2} continues the analysis with the situation after five rounds when any position in the ranking can be fixed under any of the two rules. Unsurprisingly, this has the highest probability for the group winner and the fourth-placed team. The corresponding values always exceed 15\%, and their pattern closely follows Figure~\ref{Fig1}: the first team is secured with the highest chance in Groups 1 and 4, while the last team is known most often in Group 3. The runner-up and the third-placed team are fixed only with a probability of about 10--20\%, which is reduced by at least 3.7 and at most 6.3 percentage points according to the FIFA rule.

\begin{figure}[t!]
\centering

\begin{tikzpicture}
\begin{axis}[
width = 0.48\textwidth, 
height = 0.48\textwidth,
%title = Unweighted rounds,
%title style = {font=\small},
xmajorgrids = true,
ymajorgrids = true,
symbolic x coords = {G1,G2,G3,G4},
xtick = data,
xlabel = {Groups},
xlabel style = {align=center, font=\small},
ymin = 0,
ytick distance = 1,
scaled y ticks = false,
ylabel = {Probability in \%},
ylabel style = {align=center, font=\small},
yticklabel style = {/pgf/number format/fixed,/pgf/number format/precision=5},
ytick style = {draw = none},
legend style = {font=\small,at={(1.2,0.65)},anchor=north west,legend columns=1},
legend entries = {First team (UEFA rule)$\quad$, Second team (UEFA rule), Third team (UEFA rule)$\, \, \,$,Fourth team (UEFA rule)}
]
% First team
\addplot [blue, only marks, mark = square, very thick] coordinates{
(G1,2.45955844948621)
(G2,5.59064212518124)
(G3,5.89367517419256)
(G4,2.8028376388703)
};
% Second team
\addplot [red, only marks, mark = asterisk, thick, mark size=2.5pt] coordinates{
(G1,5.25027030464924)
(G2,6.47311578213912)
(G3,7.2262730452309)
(G4,5.73593506337718)
};
% Third team
\addplot [ForestGreen, only marks, mark = triangle, very thick] coordinates{
(G1,5.57123445176126)
(G2,6.20241363236132)
(G3,5.43835151149187)
(G4,4.8317584388468)
};
% Fourth team
\addplot [brown, only marks, mark = pentagon, very thick] coordinates{
(G1,3.17082208459055)
(G2,4.80205097865207)
(G3,2.30139443383672)
(G4,5.54940846029449)
};
\end{axis}
\end{tikzpicture}

\captionsetup{justification=centering}
\caption{The probability of finishing in a different position under the FIFA rule \\ when the position is already secured under the UEFA rule before Round 6, \\ 2022/23 UEFA Nations League A}
\label{Fig3}

\end{figure}
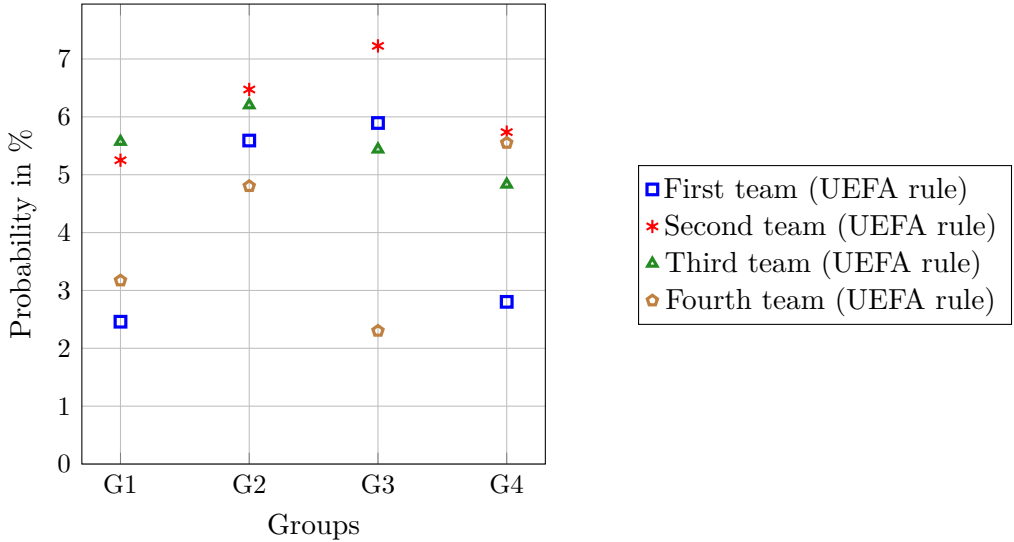

%\end{document}

As Figure~\ref{Fig3} shows, a team whose position is already secured under the UEFA rule can lose its rank if the FIFA rule is used with a chance of more than 2\%. These conditional probabilities are higher by an order of magnitude compared to the situation after four rounds, hence the teams face much more uncertainty in retaining their positions that are secured only by the UEFA rule. This is especially relevant for the two middle ranks, where another team can appear with a probability of about 5\% in the final group ranking.

For a balanced discussion of tie-breaking criteria, the arguments for the UEFA rule should also be mentioned. If goal difference is preferred, team A can be ranked over team B even if they have scored the same number of points and the head-to-head records favour team B. This might be perceived as unfair, especially if the superior goal difference of team A is (mainly) caused by scoring many goals against weaker teams (however, the design of the 2022/23 UEFA Nations League A does not allow for real underdogs).

\begin{table}[t!]
  \centering
  \caption{Potential unfairness: the probability (in \%) that the FIFA  and UEFA rules \\ rank two teams differently if there is no third team with the same number of points}
  \label{Table4}
  \rowcolors{3}{}{gray!20}
    \begin{tabularx}{0.8\textwidth}{l CCCC} \toprule
    Position & Group 1 & Group 2 & Group 3 & Group 4 \\ \bottomrule
    First--Second & 2.22  & 3.11  & 3.27  & 2.40 \\
    Second--Third & 3.01  & 3.76  & 3.36  & 3.20 \\
    Third--Fourth & 2.58  & 2.93  & 1.82  & 2.90 \\ \toprule
    \end{tabularx}
\end{table}

Therefore, Table~\ref{Table4} reports the probability that exactly two teams have the same number of points and they are ranked differently according to the FIFA and UEFA ranking rules. The values lie between 1.8\% and 3.8\% for each position in each group. Again, the difference is the largest for the two middle positions. The likelihood of such perceived unfairness is below the difference between the two ranking rules in the probability of a fixed position after five rounds. Taking into account that the UEFA rule is outperformed by the FIFA rule with respect to the chance of a secured position after four rounds (see Figure~\ref{Fig1}), the potential unfairness of the final ranking seems to be a less serious problem compared to the reduced competitiveness of the games played at the end of the tournament, where the lack of incentives to win also lead to unfairness.

To summarise, preferring goal difference to head-to-head results in tie-breaking is unambiguously beneficial for the excitement of the games. While the difference can perhaps be neglected after four rounds in a home-away round-robin tournament with four teams, the competitiveness of the last two matches is clearly increased by the FIFA rule. The effect is the strongest for the two middle teams since (1) the probability that the team is guaranteed to finish in one of these positions under the UEFA rule but not under the FIFA rule is the highest; and (2) the probability that the team in one of these positions is known under the UEFA rule but it loses its position under the FIFA rule when all group matches are played is the highest. Although the reward of the runner-up compared to the third-placed team is relatively small in the 2022/23 UEFA Nations League A, it is much higher in several prominent tournaments such as the FIFA World Cup or the UEFA Champions League, where the first two teams from each group advance to the knockout stage.

\subsection{Sensitivity analysis in a theoretical model} \label{Sec34}

In the previous section, four arbitrary sets of four teams have been analysed, which might distort the conclusions. Unfortunately, analytical results would be difficult to derive even in the case of four identical teams since the difference between the tie-breaking rules crucially depends on the number of goals scored in each match.
Thus, we have carried out simulations with a specific distribution of strengths:
\begin{itemize}
\item
Four teams play a home-away round-robin tournament;
\item
The outcomes of the matches are determined invariably by the simulation model described in Section~\ref{Sec22};
\item
There is a strong team with an Elo rating of $1900 + \Delta$;
\item
There are three weak teams with an Elo rating of $1900 - \Delta$;
\item
The values of $\{ 0; 50; 100; 150; 200 \}$ are considered for the parameter $\Delta$, which reflects the variance of strengths. 
\end{itemize}
This basic setting has been chosen to reduce the number of scheduling options. In similar tournaments, every team plays one home match and one away match in the last two rounds (see Table~\ref{Table3}), hence, two alternative orders of the games remain:
\begin{itemize}
\item
Schedule A: the strong team plays at home in Round 5 and away in Round 6;
\item
Schedule B: the strong team plays away in Round 5 and at home in Round 6.
\end{itemize}
Note also that the Elo ratings are realistic with respect to the 2022/23 UEFA Nations League, see Table~\ref{Table2}.

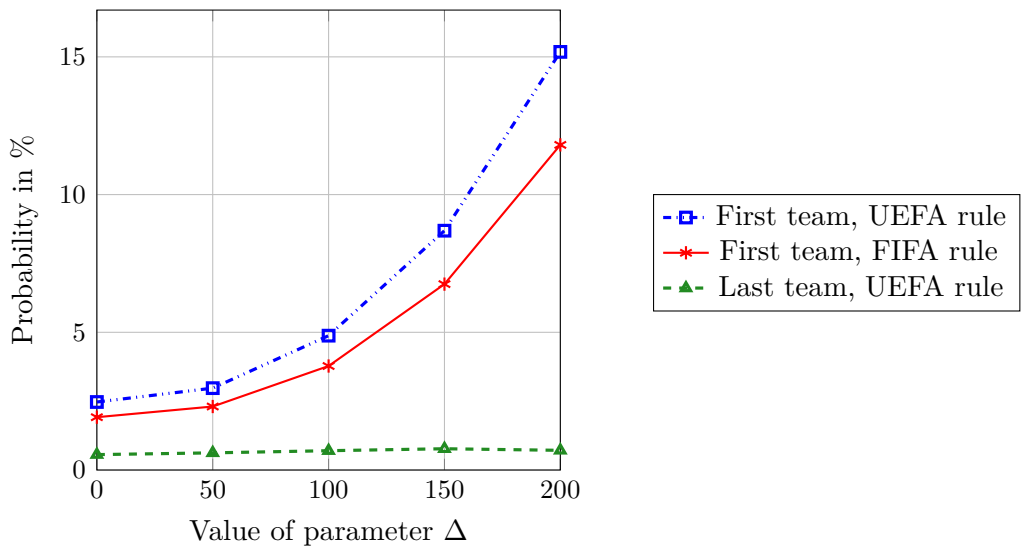
\begin{figure}[t!]
\centering

\begin{tikzpicture}
\begin{axis}[
width = 0.48\textwidth, 
height = 0.48\textwidth,
%title = Unweighted rounds,
%title style = {font=\small},
xmajorgrids = true,
ymajorgrids = true,
%symbolic x coords = {G1,G2,G3,G4},
%xtick = data,
xlabel = {Value of parameter $\Delta$},
xlabel style = {align=center, font=\small},
xmin = 0,
xmax = 200,
ymin = 0,
%ymode = log,
ylabel = {Probability in \%},
ylabel style = {align=center, font=\small},
%log ticks with fixed point,
yticklabel style = {/pgf/number format/fixed,/pgf/number format/precision=3},
ytick style = {draw = none},
legend style = {font=\small,at={(1.2,0.6)},anchor=north west,legend columns=1},
legend entries = {{First team, UEFA rule}, {First team, FIFA rule}$\,$, {Last team, UEFA rule}}
]
% First team UEFA, Schedule A
\addplot [blue, dashdotdotted, mark = square, mark options = solid, very thick] coordinates{
(0,2.4696)
(50,2.9729)
(100,4.8781)
(150,8.6888)
(200,15.1807)
};
% First team FIFA, Schedule A
\addplot [red, mark = asterisk, mark options = solid, thick, mark size=2.5pt] coordinates{
(0,1.9177)
(50,2.3056)
(100,3.776)
(150,6.7493)
(200,11.8044)
};
% Last team UEFA, Schedule A
\addplot [ForestGreen, dashed, mark options = solid, mark = triangle, very thick] coordinates{
(0,0.5594)
(50,0.6233)
(100,0.7015)
(150,0.7724)
(200,0.7148)
};
\end{axis}
\end{tikzpicture}

\captionsetup{justification=centering}
\caption{The probability of an already secured position before Round 5, theoretical model}
\label{Fig4}

\end{figure}

%\end{document}

The probability of a fixed position after four matchdays is shown in Figure~\ref{Fig4}. Compared to the FIFA rule, favouring head-to-head records means that the first and the last team are known with a higher probability. The increase is about 0.5 percentage points for both positions even in the case of identical teams ($\Delta = 0$). For the group winner, the difference gradually increases to exceed 3 percentage points if there is a dominant team in the group. On the other hand, the probability that relegation is decided after four rounds stabilises close to 75 basis points (0.75\%) due to the presence of three weak teams. Since the strong team should play one match at home and one away in both Schedules A and B, the order of the games does not affect these values.

\begin{figure}[t!]
\centering

\begin{tikzpicture}
%\selectcolormodel{gray}
\begin{axis}[
name = axis1,
width = 0.48\textwidth, 
height = 0.48\textwidth,
title = {Group winner},
title style = {align=center, font=\small},
xmajorgrids = true,
ymajorgrids = true,
%symbolic x coords = {G1,G2,G3,G4},
%xtick = data,
xlabel = {Value of parameter $\Delta$},
xlabel style = {align=center, font=\small},
xmin = 0,
xmax = 210,
ymin = 0,
scaled y ticks = false,
ylabel = {Probability in \%},
ylabel style = {align=center, font=\small},
%log ticks with fixed point,
yticklabel style = {/pgf/number format/fixed,/pgf/number format/precision=5},
ytick style = {draw = none},
]
% UEFA rule, Schedule A
\addplot [blue, dashdotdotted, mark = square, mark options = solid, very thick] coordinates{
(0,17.0465)
(50,25.3166)
(100,40.0945)
(150,56.9344)
(200,75.6156)
};
% FIFA rule, Schedule A
\addplot [red, mark = asterisk, mark options = solid, thick, mark size=2.5pt] coordinates{
(0,12.8677)
(50,20.248)
(100,34.049)
(150,50.5605)
(200,69.847)
};
% UEFA rule, Schedule B
\addplot [ForestGreen, dashed, mark = triangle, mark options = solid, very thick] coordinates{
(0,16.7542)
(50,19.3266)
(100,30.3295)
(150,46.4829)
(200,65.6383)
};
% FIFA rule, Schedule B
\addplot [brown, dotted, mark = pentagon, mark options = solid, very thick] coordinates{
(0,12.5644)
(50,14.8993)
(100,25.1053)
(150,40.6263)
(200,59.9916)
};
\end{axis}

\begin{axis}[
at = {(axis1.south east)},
xshift = 0.15\textwidth,
width = 0.48\textwidth, 
height = 0.48\textwidth,
title = {Runner-up},
title style = {align=center, font=\small},
xmajorgrids = true,
ymajorgrids = true,
%symbolic x coords = {G1,G2,G3,G4},
%xtick = data,
xlabel = {Value of parameter $\Delta$},
xlabel style = {align=center, font=\small},
xmin = 0,
xmax = 210,
ymin = 0,
%ymax = 26,
%ytick distance = 5,
scaled y ticks = false,
ylabel = {Probability in \%},
ylabel style = {align=center, font=\small},
yticklabel style = {/pgf/number format/fixed,/pgf/number format/precision=5},
ytick style = {draw = none},
]
% UEFA rule, Schedule A
\addplot [blue, dashdotdotted, mark = square, mark options = solid, very thick] coordinates{
(0,16.7289)
(50,17.662)
(100,18.5778)
(150,19.5932)
(200,20.3076)
};
% FIFA rule, Schedule A
\addplot [red, mark = asterisk, mark options = solid, thick, mark size=2.5pt] coordinates{
(0,12.5828)
(50,13.342)
(100,14.0254)
(150,14.7991)
(200,15.237)
};
% UEFA rule, Schedule B
\addplot [ForestGreen, dashed, mark = triangle, mark options = solid, very thick] coordinates{
(0,17.0114)
(50,16.0993)
(100,15.9024)
(150,16.4389)
(200,17.5634)
};
% FIFA rule, Schedule B
\addplot [brown, dotted, mark = pentagon, mark options = solid, very thick] coordinates{
(0,12.8305)
(50,12.0237)
(100,11.7121)
(150,11.8382)
(200,12.3952)
};
\end{axis}
\end{tikzpicture}

\vspace{0.2cm}
\begin{tikzpicture}
\begin{axis}[
name = axis3,
width = 0.48\textwidth, 
height = 0.48\textwidth,
title = {Third-placed team},
title style = {align=center, font=\small},
xmajorgrids = true,
ymajorgrids = true,
%symbolic x coords = {G1,G2,G3,G4},
%xtick = data,
xlabel = {Value of parameter $\Delta$},
xlabel style = {align=center, font=\small},
xmin = 0,
xmax = 210,
ymin = 0,
%ymax = 26,
%ytick distance = 5,
%scaled y ticks = false,
ylabel = {Probability in \%},
ylabel style = {align=center, font=\small},
yticklabel style = {/pgf/number format/fixed,/pgf/number format/precision=5},
ytick style = {draw = none},
legend style = {font=\small,at={(0.3,-0.25)},anchor=north west,legend columns=2},
legend entries = {{UEFA rule, Schedule A}$\qquad \qquad$, {FIFA rule, Schedule A},{UEFA rule, Schedule B}$\qquad \qquad$, {FIFA rule, Schedule B}}
]
% UEFA rule, Schedule A
\addplot [blue, dashdotdotted, mark = square, mark options = solid, very thick] coordinates{
(0,16.7664)
(50,17.6501)
(100,19.0109)
(150,20.936)
(200,23.8237)
};
% FIFA rule, Schedule A
\addplot [red, mark = asterisk, mark options = solid, thick, mark size=2.5pt] coordinates{
(0,12.5601)
(50,12.7422)
(100,12.8693)
(150,13.378)
(200,14.4806)
};
% UEFA rule, Schedule B
\addplot [ForestGreen, dashed, mark = triangle, mark options = solid, very thick] coordinates{
(0,17.0753)
(50,16.488)
(100,16.7282)
(150,18.05)
(200,20.6342)
};
% FIFA rule, Schedule B
\addplot [brown, dotted, mark = pentagon, mark options = solid, very thick] coordinates{
(0,12.8952)
(50,12.3575)
(100,11.9523)
(150,12.1055)
(200,13.0314)
};
\end{axis}

\begin{axis}[
at = {(axis3.south east)},
xshift = 0.15\textwidth,
width = 0.48\textwidth, 
height = 0.48\textwidth,
title = {Fourth-placed team},
title style = {align=center, font=\small},
xmajorgrids = true,
ymajorgrids = true,
%symbolic x coords = {G1,G2,G3,G4},
%xtick = data,
xlabel = {Value of parameter $\Delta$},
xlabel style = {align=center, font=\small},
xmin = 0,
xmax = 210,
ymin = 0,
%ymax = 55,
%ytick distance = 10,
scaled y ticks = false,
ylabel = {Probability in \%},
ylabel style = {align=center, font=\small},
yticklabel style = {/pgf/number format/fixed,/pgf/number format/precision=5},
ytick style = {draw = none},
]
% UEFA rule, Schedule A
\addplot [blue, dashdotdotted, mark = square, mark options = solid, very thick] coordinates{
(0,16.7664)
(50,17.6501)
(100,19.0109)
(150,20.936)
(200,23.8237)
};
% FIFA rule, Schedule A
\addplot [red, mark = asterisk, mark options = solid, thick, mark size=2.5pt] coordinates{
(0,12.5601)
(50,12.7422)
(100,12.8693)
(150,13.378)
(200,14.4806)
};
% UEFA rule, Schedule B
\addplot [ForestGreen, dashed, mark = triangle, mark options = solid, very thick] coordinates{
(0,16.7147)
(50,17.8968)
(100,19.0633)
(150,19.8934)
(200,20.6785)
};
% FIFA rule, Schedule B
\addplot [brown, dotted, mark = pentagon, mark options = solid, very thick] coordinates{
(0,12.5314)
(50,13.8859)
(100,15.0843)
(150,15.8113)
(200,16.2874)
};
\end{axis}
\end{tikzpicture}

\caption{The probability of an already secured position before Round 6, theoretical model}
\label{Fig5}

\end{figure}
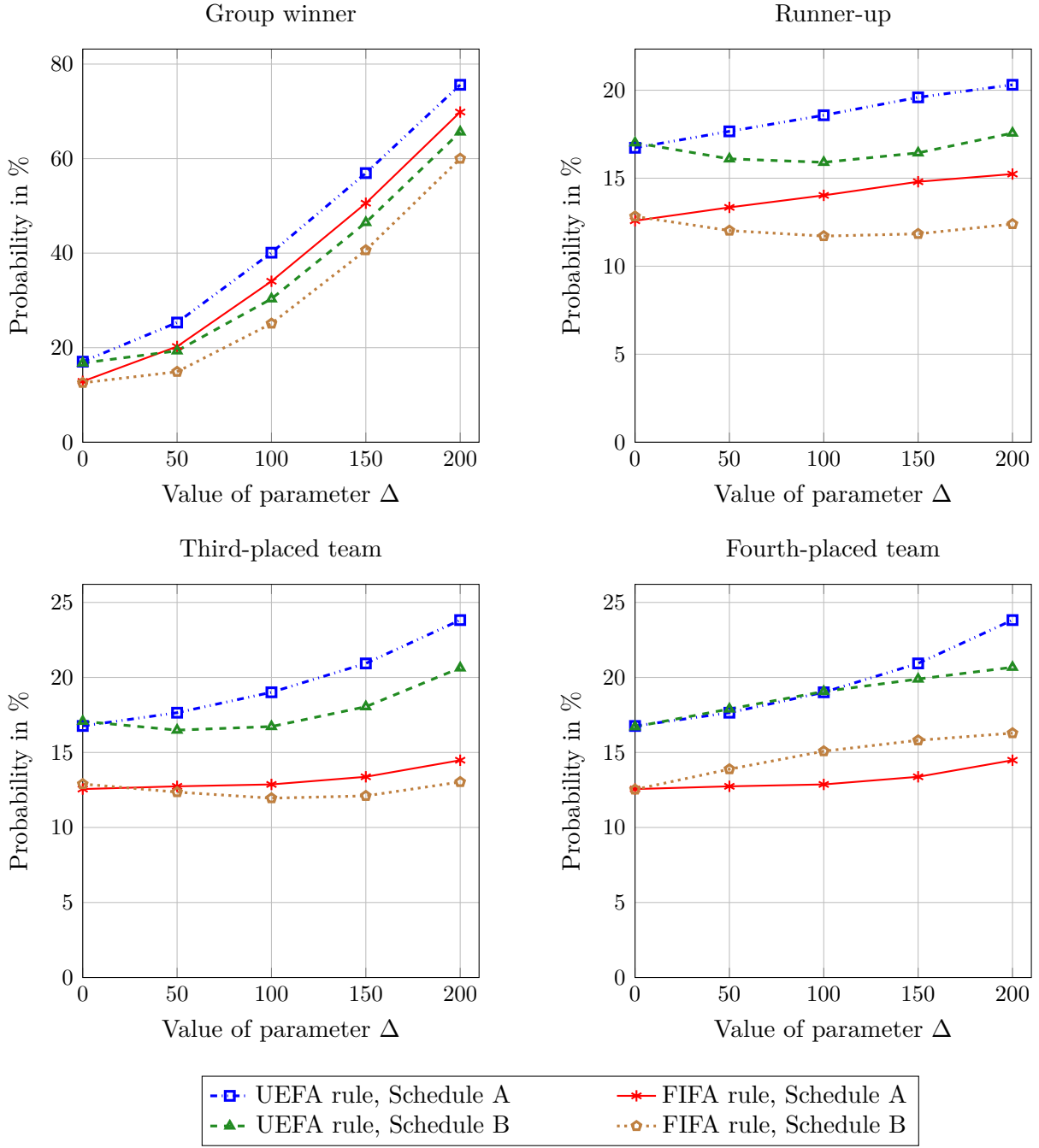

%\end{document}

On the other hand, as Figure~\ref{Fig5} uncovers, scheduling does strongly influence the chance that the position of a team will be secured after five rounds. In particular, the probability of a known group winner can be higher by 10 percentage points if the strong team plays away in the last round (Schedule A) and the group is imbalanced ($\Delta \geq 100$). More importantly, the FIFA ranking rule has a robust advantage of 5 percentage points over the UEFA rule from this point of view, which is quite substantial in relative terms, corresponding to a reduction of 20--25\%. The effect of the schedule is more mitigated for the other three positions, however, the improvement caused by preferring goal difference in tie-breaking does not decrease below 5 percentage points. The two schedules are identical if $\Delta = 0$, the small differences are owing to the stochastic nature of the simulation. The likelihood of an already secured position is generally higher if the variance of strengths is increased.

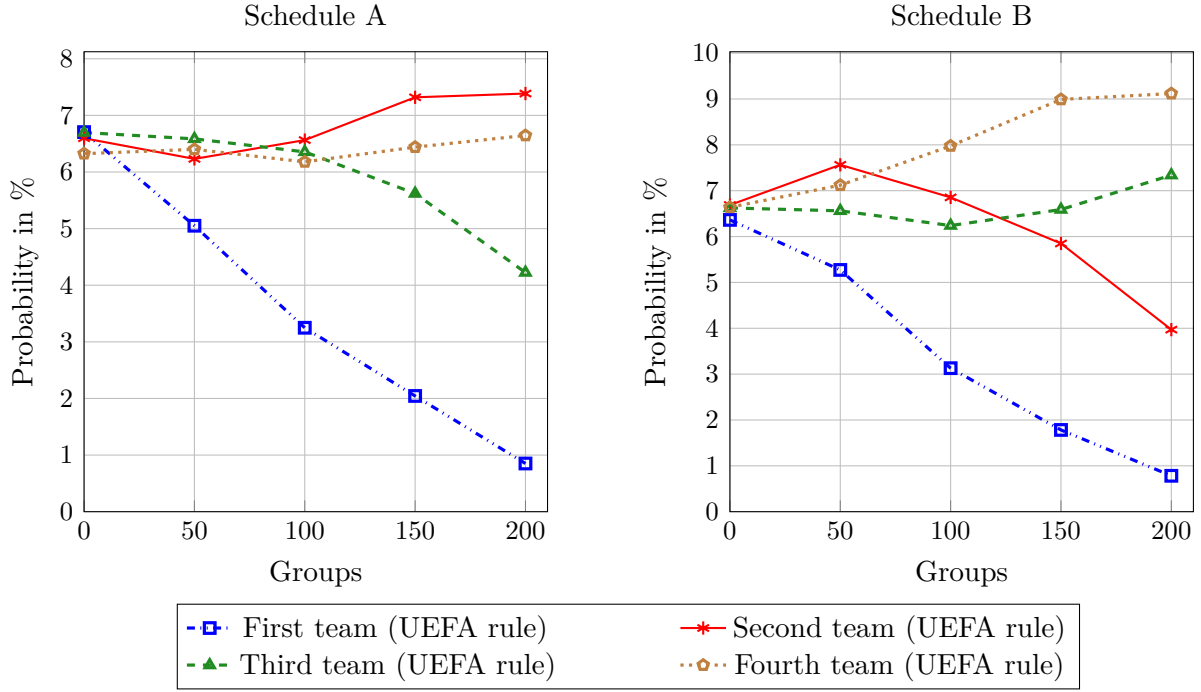
\begin{figure}[t!]
\centering

\begin{tikzpicture}
\begin{axis}[
name = axis1,
width = 0.48\textwidth, 
height = 0.48\textwidth,
title = Schedule A,
title style = {font=\small},
xmajorgrids = true,
ymajorgrids = true,
%symbolic x coords = {G1,G2,G3,G4},
%xtick = data,
xlabel = {Groups},
xlabel style = {align=center, font=\small},
xmin = 0,
xmax = 210,
ymin = 0,
ytick distance = 1,
scaled y ticks = false,
ylabel = {Probability in \%},
ylabel style = {align=center, font=\small},
yticklabel style = {/pgf/number format/fixed,/pgf/number format/precision=5},
ytick style = {draw = none},
legend style = {font=\small,at={(0.2,-0.2)},anchor=north west,legend columns=2},
legend entries = {First team (UEFA rule)$\, \qquad \qquad$,Second team (UEFA rule),Third team (UEFA rule)$\qquad \qquad$,Fourth team (UEFA rule)}
]
% First team
\addplot [blue, dashdotdotted, mark = square, mark options = solid, very thick] coordinates{
(0,6.70527424140902)
(50,5.05070433650318)
(100,3.2486973782152)
(150,2.04427430615479)
(200,0.851159726796797)
};
% Second team
\addplot [red, mark = asterisk, mark options = solid, thick, mark size=2.5pt] coordinates{
(0,6.59656062323629)
(50,6.23148148148148)
(100,6.56357086372024)
(150,7.31941344569367)
(200,7.38571372224194)
};
% Third team
\addplot [ForestGreen, dashed, mark = triangle, mark options = solid, very thick] coordinates{
(0,6.69698123988192)
(50,6.58523873189966)
(100,6.35553810222599)
(150,5.61962636163528)
(200,4.22421198704817)
};
% Fourth team
\addplot [brown, dotted, mark = pentagon, mark options = solid, very thick] coordinates{
(0,6.32384756199035)
(50,6.39988589824569)
(100,6.17917155138726)
(150,6.43953426832495)
(200,6.64554591088611)
};
\end{axis}

\begin{axis}[
at = {(axis1.south east)},
xshift = 0.15\textwidth,
width = 0.48\textwidth, 
height = 0.48\textwidth,
title = Schedule B,
title style = {font=\small},
xmajorgrids = true,
ymajorgrids = true,
%symbolic x coords = {G1,G2,G3,G4},
%xtick = data,
xlabel = {Groups},
xlabel style = {align=center, font=\small},
xmin = 0,
xmax = 210,
ymin = 0,
ytick distance = 1,
scaled y ticks = false,
ylabel = {Probability in \%},
ylabel style = {align=center, font=\small},
yticklabel style = {/pgf/number format/fixed,/pgf/number format/precision=5},
ytick style = {draw = none},
]
% First team
\addplot [blue, dashdotdotted, mark = square, mark options = solid, very thick] coordinates{
(0,6.36307222301781)
(50,5.27183610778578)
(100,3.12966578614908)
(150,1.78260424136871)
(200,0.782758071085767)
};
% Second team
\addplot [red, mark = asterisk, mark options = solid, thick, mark size=2.5pt] coordinates{
(0,6.68755531105743)
(50,7.56453037589557)
(100,6.85392454000907)
(150,5.84910991805595)
(200,3.97043458070508)
};
% Third team
\addplot [ForestGreen, dashed, mark = triangle, mark options = solid, very thick] coordinates{
(0,6.62424343915217)
(50,6.56094903764677)
(100,6.23966163445633)
(150,6.58928421229708)
(200,7.33940127321513)
};
% Fourth team
\addplot [brown, dotted, mark = pentagon, mark options = solid, very thick] coordinates{
(0,6.63829990677217)
(50,7.12059637487846)
(100,7.97687861271676)
(150,8.99047059111732)
(200,9.11844412561773)
};
\end{axis}
\end{tikzpicture}

\captionsetup{justification=centering}
\caption{The probability of finishing in a different position under the FIFA rule when the position is already secured under the UEFA rule before Round 6, theoretical model}
\label{Fig6}

\end{figure}

%\end{document}

Finally, Figure~\ref{Fig6} presents the probability that a team with a known position under the UEFA rule loses its rank if goal difference is preferred to break the ties. In the case of four identical teams, this lies above 6\%, which is much higher compared to the real-world study in Figure~\ref{Fig3}. While this conditional probability rapidly decreases for the group winner as the parameter $\Delta$ grows, the teams competing for the other three places face much uncertainty in holding their position which would be secured only by favouring head-to-head results. Consequently, the advantage of the FIFA rule in the competitiveness of the games seems to be more pronounced when the teams are closer in strength.

Our theoretical investigation has reinforced the findings from the simulations based on the 2022/23 UEFA Nations League. In particular, preferring goal difference to head-to-head records is especially useful to increase the stakes of the games played in the last two rounds if the competition is balanced because no team can be calm to be ranked over another merely due to some luck in the already played matches. Therefore, besides the widely known role of scheduling \citep{ChaterArrondelGayantLaslier2021, Guyon2020a, Stronka2020}, tie-breaking criteria need to be considered as another crucial aspect of fair tournament design.

\section{Conclusions} \label{Sec4}

This paper has analysed two popular tie-breaking concepts in round-robin contests from a novel perspective, focusing on their implications for the competitiveness of the games played in the last round(s). A real-world example has revealed that a team could be guaranteed to win a round-robin tournament if head-to-head results are considered over goal difference but not if the latter indicator is preferred. According to simulations based on the 2022/23 UEFA Nations League A, the difference between the two basic tie-breaking principles---used by the FIFA and UEFA, among others---is non-negligible. The priority of head-to-head records makes the position of the middle teams less uncertain, thus it can be detrimental to attendance especially if the first two teams qualify from a group of four, which is the case in several prominent tournaments.
Based on the calculations above, it is hard to argue for favouring head-to-head results over goal difference to break the ties.

Our finding yields an important lesson for tournament organisers by highlighting that the seemingly innocent order of tie-breaking criteria may have fundamental sporting effects. While previous studies have already explored the attractiveness of giving priority to goal difference instead of head-to-head results, as well as the crucial role of scheduling to avoid match-fixing opportunities, the latter issue has been verified here to be an essential aspect of determining ranking systems. Consequently, tie-breaking rules are worth getting more attention in the economic design of sporting contests.

\section*{Acknowledgements}
\addcontentsline{toc}{section}{Acknowledgements}
\noindent
This paper could not have been written without \emph{my father} (also called \emph{L\'aszl\'o Csat\'o}), who has primarily coded the simulations in Python. \\
%We are grateful to \emph{Tam\'as Halm} and \emph{D\'ora Gr\'eta Petr\'oczy} for useful advice. \\
Three anonymous reviewers provided valuable comments and suggestions on an earlier draft. \\
We are indebted to the \href{https://en.wikipedia.org/wiki/Wikipedia_community}{Wikipedia community} for summarising important details of the sports competition discussed in the paper. \\
The research was supported by the MTA Premium Postdoctoral Research Program grant PPD2019-9/2019.

\bibliographystyle{apalike}
\bibliography{All_references}

\end{document}